\documentstyle[preprint,version2,aps]{revtex}
\begin{document}
\draft
\begin{title}
N\'eel temperature for quasi two-dimensional dipolar antiferromagnets
\end{title}
\author{C. Pich and F. Schwabl}
\begin{instit}
Institut f\"ur Theoretische Physik, \\
Physik-Department der Technischen Universit\"at M\"unchen, \\
D-85747 Garching, Federal Republic of Germany
\end{instit}
\begin{abstract}
We calculate the  N\'eel temperature $T_N$ for two-dimensional isotropic
dipolar Heisenberg antiferromagnets via linear spin-wave theory and a high
temperature expansion, employing the method of Callen. The theoretical
predictions for $T_N$ for K$_2$MnF$_4$, Rb$_2$MnF$_4$, Rb$_2$MnCl$_4$ and
(CH$_3$NH$_3$)$_2$MnCl$_4$ are in good agreement with the measured values.
\end{abstract}
\pacs{PACS numbers: 75.10 J, 75.30 D, 75.30 G, 75.50 E}
\narrowtext

Recently\cite{pich} it has been shown that long-range order is possible in
two-dimensional isotropic Heisenberg antiferromagnets due to the anisotropy
of the dipole-dipole interaction. The occurence of a finite energy gap goes in
hand with a nonvanishing order parameter and a finite N\'eel-temperature.
In the present paper a quantitative improvement of the theory is achieved by
means of Callen's extension of the Tyablikov decoupling scheme \cite{Cal}.

In\cite{pich} we have used linear spin wave theory based on the
Holstein-Primakoff transformation to evaluate the magnon dispersion relation.
The evaluation of the N\'eel temperature by means of the temperature
independent dispersion relation leads to an overestimate of $T_N$. In reality
the magnon frequency softens with increasing temperature and thus the actual
transition temperature is lower. This feature is accounted for by an extension
of the Tyablikov decoupling scheme due to Callen\cite{Cal}. In essence the
dependence on the magnitude of the spin $S$ is replaced by $\sigma$, the
temperature dependent order parameter. Even at $T=0$ the zero point
fluctuations lead to a reduction of $\sigma$ as compared to $S$.
The resulting transition temperature is lowered in comparison to the estimate
of Ref.\cite{pich} such that a satisfactory agreement between theory and
experiment is achieved.

The Hamiltonian of a dipolar antiferromagnet reads
\begin{equation}
H=-\sum_{l\ne l'}\sum_{\alpha\beta}\left(J_{ll'}\delta_{\alpha\beta}+
A^{\alpha\beta}_{ll'}\right)S_l^{\alpha}S_{l'}^{\beta}~,
\end{equation}
with spins ${\bf S}_l$ at lattice sites ${\bf x}_l$. The first term in brackets
is the exchange interaction $J_{ll'}$ and the second the usual dipole-dipole
interaction.
We consider a square lattice in the $xy$ plane with lattice constant $a$ and
the spins orientated alternatingly along the $z$ axis. The out-of-plane
orientation is the classical ground state for the isotropic dipolar
antiferromagnet with a nearest-neighbor exchange energy $|J|$ much larger than
the dipole energy \cite{pich}.

Let us introduce the retarded double-time Green functions according to Callen
\cite{Cal}
\begin{mathletters}
\begin{eqnarray}
G^1({\bf R}_k-{\bf R}_0,t)&=&-i\Theta (t)\langle[S^+_k(t),e^{bS^z_0}S^-_0(0)]
\rangle \equiv \langle\langle S^+_k|e^{bS^z_0}S^-_0\rangle\rangle, \\
G^2({\bf R}_k-{\bf R}_0,t)&=&-i\Theta (t)\langle[S^-_k(t),e^{bS^z_0}S^-_0(0)]
\rangle \equiv \langle\langle S^-_k|e^{bS^z_0}S^-_0\rangle\rangle ,
\end{eqnarray}
\end{mathletters}
which obey the following Fourier transformed equation of motion\cite{elk,Zub}
\begin{mathletters}
\begin{eqnarray}
\omega\langle\langle S^+_k|e^{bS^z_0}S^-_0 \rangle\rangle_\omega &=& \Theta_b
{}~+~\langle\langle [S^+_k,H]|e^{bS^z_0}S^-_0\rangle\rangle_\omega ~,\\
\omega\langle\langle S^-_k|e^{bS^z_0}S^-_0 \rangle\rangle_\omega &=&
\langle\langle [S^-_k,H]|e^{bS^z_0}S^-_0\rangle\rangle_\omega ~,
\end{eqnarray}
\end{mathletters}
with the equal time commutator
\begin{eqnarray}
\Theta_b & = & \langle[S^+,e^{bS^z}S^-] \rangle.\nonumber
\end{eqnarray}
Here ${\bf R}_0$ is a lattice point of sublattice one and ${\bf R}_k$ is a
general lattice vector. The higher order Green functions generated by the
commutator in Eqs. (3a) and (3b) are approximated by the Tyablikov decoupling
scheme
\begin{equation}
\langle\langle S^z_lS^+_m|e^{bS^z_0}S^-_0\rangle\rangle ~~ \to ~~
\langle S^z_l \rangle \langle\langle S^+_m|e^{bS^z_0}S^-_0\rangle\rangle .
\end{equation}
As a consequence of translational symmetry the mean spin value is independent
of the lattice site for each sublattice: $ \langle S^z_{l_1}\rangle = -
\langle S^z_{l_2}\rangle =\sigma$ where $l_1\in L_1$ ($l_2 \in L_2$) refers to
sublattice one (two). Using this approximation we obtain from Eqs. (3a) and Eq.
(3b) a set of four equations for the Green functions $G^1_{\bf q}$, $G^1_{{\bf
q}+{\bf q}_0}$, $G^2_{\bf q}$ and $G^2_{{\bf q}+{\bf q}_0}$. The evaluation of
the order parameter and the magnon dispersion relation requires only two of
them:
\begin{eqnarray}
\tilde G^1_{\bf q}(\omega) \equiv {1\over 2}(G^1_{\bf q}(\omega)+
G^1_{{\bf q}+{\bf q}_0}(\omega)) &=&\Theta_b {A^{(1)}+B^{(1)}\omega
+C^{(1)}\omega^2 +\omega^3\over
(\omega^2-\epsilon_1^2)(\omega^2-\epsilon_2^2)},
\end{eqnarray}
with
\begin{eqnarray}
-A^{(1)} & = &  {1\over 2}A_{\bf q}(A_{{\bf q}+{\bf q}_0}^2-B_{{\bf q}+
{\bf q}_0}^2)+{1\over 2}A_{{\bf q}+{\bf q}_0}(A_{\bf q}^2-B_{\bf q}^2)
\nonumber \\
&& \quad +(C_{\bf q}^2+C_{{\bf q}+{\bf q}_0}^2)(A_{\bf q}+A_{{\bf q}+
{\bf q}_0})+(C_{\bf q}^2-C_{{\bf q}+{\bf q}_0}^2)(B_{{\bf q}+{\bf q}_0}-
B_{\bf q}),
\nonumber\\
-B^{(1)} & = & {1\over 2}( A_{\bf q}^2-B_{\bf q}^2+A_{{\bf q}+{\bf q}_0}^2-
B_{{\bf q}+{\bf q}_0}^2+8C_{\bf q}C_{{\bf q}+{\bf q}_0}),
\nonumber\\
C^{(1)} & = & {1\over 2}(A_{\bf q}+A_{{\bf q}+{\bf q}_0}).\nonumber
\end{eqnarray}
Here ${\bf q}$ denotes the wave-vector of the chemical Brillouin zone and
${\bf q}_0={\pi\over a}(1,1,0)$. In the magnetic Brillouin zone which is
half the chemical there exist two distinct spin-wave branches with frequencies
$\epsilon_i $ ($i=1,2$) which read
\begin{equation}
\epsilon_i^2 = {1\over 2}(\Omega_1 \pm \Omega_2),
\end{equation}
\[
\Omega_1 =A_{\bf q}^2-B_{\bf q}^2+A_{{\bf q}+{\bf q}_0}^2-B_{{\bf q}+{\bf
q}_0}^2 +8C_{\bf q}~C_{{\bf q}+{\bf q}_0}~,
\]
\[\Omega_2^2 = (A_{\bf q}^2-B_{\bf q}^2-A_{{\bf q}+{\bf q}_0}^2+B_{{\bf q}+
{\bf q}_0}^2)^2+
16[C_{{\bf q}+{\bf q}_0}(A_{{\bf q}+{\bf q}_0}-B_{{\bf q}+{\bf q}_0})-
C_{\bf q}~ (A_{\bf q}-B_{\bf q})]
\]
\[ \times [C_{\bf q}~(A_{{\bf q}+{\bf q}_0}+B_{{\bf q}+{\bf q}_0})-
C_{{\bf q}+{\bf q}_0}(A_{\bf q}+B_{\bf q})].\]
In Eq. (6) the coefficients
\begin{mathletters}
\begin{eqnarray}
A_{\bf q}  & = & \sigma(2J_{\bf q_0}-J_{\bf q}-J_{{\bf q}+{\bf q}_0})+
\sigma(2A^{zz}_{{\bf q}_0}-A^{xx}_{\bf q}-A^{yy}_{{\bf q}+{\bf q}_0}),\\
B_{\bf q}  & = & \sigma(J_{{\bf q}+{\bf q}_0}-J_{\bf q}) +
\sigma(A^{yy}_{{\bf q}+{\bf q}_0}-A^{xx}_{\bf q}),\\
C_{\bf q}  & = &  i\sigma A^{xy}_{\bf q},
\end{eqnarray}
\end{mathletters}
have been introduced. This result, Eq. (6), coincides  with the magnon
frequencies derived by the Holstein-Primakoff transformation when $\sigma$ is
replaced by $S$ and the external magnetic field is set to zero\cite{pich}. But
there is a difference even at absolute zero
temperature, because the ground state of the antiferromagnet is not the N\'eel
state of fully aligned spins, i.e. $\sigma (T=0) < S$ as will be seen later.
Note that the magnon frequency scales with the order parameter $\sigma$, i.e.
the whole spectrum softens with increasing temperatures and vanishes at the
phase transition.

Now we turn to the evaluation of $\sigma$. For arbitrary spin $S$ the spin
expectation value is given by the well known relation
\begin{eqnarray}
\langle S^z \rangle = S(S+1)-\langle(S^z)^2 \rangle -\langle S^-S^+ \rangle.
\end{eqnarray}
For $S=1/2$ the order parameter can be calculated directly via the Green
functions of Eq. (2a) and (2b) with $b\equiv 0$ and $\Theta_0=2\sigma$. For
higher spin quantum numbers the above formula is not so helpful. Then
a convenient starting point is the following gerneralized thermal
average\cite{Cal}
\begin{eqnarray}
\psi (b)\equiv \langle e^{bS_0^z}S_0^-S_0^+ \rangle,
\end{eqnarray}
by means of which a self-consistent system of equations can be derived by the
method of Callen. The above thermal average can be represented by the spectral
theorem\cite{elk,Zub}:
\begin{eqnarray}
\langle e^{bS_0^z}S_0^-S_0^+ \rangle & = & \lim_{\delta\to 0}-{i\over 2\pi}
\int_{-\infty}^\infty d\omega \ n(\omega){1\over N}\sum_{\bf q}
\{\tilde G^1_{\bf q}(\omega+i\delta)-\tilde G^1_{\bf
q}(\omega-i\delta)\}\nonumber \\
&&\nonumber\\
&=& \Theta_b (n-1/2),
\end{eqnarray}
where Eq. (5) has been used in the last step and $N$ denotes the total number
of spins. Here we have introduced the Bose occupation number
\begin{eqnarray}
n(\omega) = \left(e^{\omega\over k_BT}-1\right)^{-1}~,\nonumber
\end{eqnarray}
and
\begin{eqnarray}
n &=& {1\over N} \sum_{\bf q}{-A^{(1)}+C^{(1)}\epsilon_1\epsilon_2 \over 2
\epsilon_1\epsilon_2(\epsilon_1+\epsilon_2)}+{A^{(1)}+C^{(1)}
\epsilon_1^2\over\epsilon_1(\epsilon_1^2-\epsilon_2^2)}n(\epsilon_1)+{A^{(1)}+
C^{(1)}\epsilon_2^2\over \epsilon_2(\epsilon_2^2-\epsilon_1^2)}n(\epsilon_2)~.
\end{eqnarray}
The right hand side of Eq. (11) depends on $\sigma$ only via the occupation
numbers $n(\epsilon_i)$. For spin $S=1/2$ and vanishing parameter $b$ the
thermal  average [Eq. (9)] represents the number of spin wave excitations,
which reduce the staggered magnetization from the totally ordered N\'eel state,
not only for finite but also for zero temperature.

For arbitrary spin one has to express $\sigma$ in terms of $n$ which can be
achieved by the method of Callen with the result\cite{Cal,Yab91}:
\begin{equation}
\sigma=(S+{1\over 2}){(n+1/2)^{2S+1}+(n-1/2)^{2S+1}\over
(n+1/2)^{2S+1}-(n-1/2)^{2S+1}}-n.
\end{equation}
Eqs. (11) and (12) constitute a self-consistent system of equations for $n$ and
the spin expectation value $\sigma$.

Let us now discuss the dispersion relation for $T=0$. In this limit Eq. (11)
reduces to
\begin{eqnarray}
n_0\equiv n(T=0) = {1\over N} \sum_{\bf q}{-A^{(1)}+C^{(1)}\epsilon_1\epsilon_2
\over 2 \epsilon_1\epsilon_2(\epsilon_1+\epsilon_2)},
\end{eqnarray}
where the right-hand side is independent of $\sigma$. Let us denote the spin
expectation value at zero temperature by $\sigma_0=\sigma (T=0)$ which is found
by inserting Eq. (13) into (12). Knowing $\sigma_0$ one obtains for the
staggered magnetization $N(0)$ at $T=0$:
\begin{eqnarray}
N(0)=g\mu_BN\sigma_0.
\end{eqnarray}
One can convince oneself that Eq. (14) for large $S$ coincides with the
expression derived by the Holstein-Primakoff transformation\cite{pich} (see
also \cite{Yab91}). This must be so because the latter is an expansion
in $1/S$. In addition we derive an energy gap ({\bf q} $=0$) from Eq. (6)
\begin{equation}
E_0^\sigma \equiv \epsilon_1(T=0)=\epsilon_2(T=0)= 2\sigma_0\sqrt{A^{zz}_{\bf
q_0}-A^{\rho\rho}_{{\bf q}_0}}\sqrt{(J_{{\bf q}_0}-J_0)-
(A^{\rho\rho}_0-A^{zz}_{{\bf q}_0})}~,
\end{equation}
with
\begin{eqnarray}
A^{\rho\rho}_0=A^{xx}_0=A^{yy}_0,\qquad A^{\rho\rho}_{{\bf q}_0}= A^{xx}_{{\bf
q}_0}=A^{yy}_{{\bf q}_0}.\nonumber
\end{eqnarray}
This is of the same form as the result from the Holstein-Primakoff
transformation $E_0$\cite{pich} except for the prefactor which is smaller
by the ratio $\sigma_0 /S$.

Now we turn to the evaluation of the transition temperature $T_N$, i.e.
consider the limit $\sigma\to 0$.
Since the spin-wave energy (Eq. 6) is proportional to $\sigma$ the Bose
occupation numbers can be replaced by their classical limit:
\begin{eqnarray}
n(\epsilon_i)\to {k_BT\over \epsilon_i}~~. \nonumber
\end{eqnarray}
If this is inserted into Eq. (11) together with $\sigma\to 0$ one obtains
\begin{eqnarray}
n = {k_BT_N\over N\sigma} \sum_{\bf q}{-\tilde A^{(1)}\over
(\tilde\epsilon_1\tilde\epsilon_2)^2}~~.
\end{eqnarray}
To keep track of the $\sigma$ dependence we have introduced the
$\sigma$-independent quantities $\tilde A^{(1)}=A^{(1)}/\sigma^3$ and
$\tilde\epsilon_i = \epsilon_i/\sigma $. According to Eq. (16) $n$ increases
indefinitly with $\sigma\to 0$ and thus the second relation between $\sigma$
and $n$, Eq. (12), becomes\cite{Cal}
\begin{equation}
\sigma = {S(S+1)\over 3}{1\over n}.
\end{equation}
Combining Eqs. (16) and (17) we obtain an explicit expression for the N\'eel
temperature:
\begin{eqnarray}
T_N &=& {S(S+1)\over 3k_B}F^{-1}
\end{eqnarray}
with
\begin{eqnarray}
F &=& {1\over N} \sum_{\bf q}{-\tilde A^{(1)}\over
(\tilde\epsilon_1\tilde\epsilon_2)^2}~~.\nonumber
\end{eqnarray}
For purely isotropic antiferromagnets the coefficient $F$ diverges,
excluding long-range order at finite temperature in two dimensions in accord
with the Hohenberg-Mermin-Wagner theorem\cite{Hohen67,Mermin66}.

In the presence of the dipolar interaction there is an energy gap and $T_N$
becomes finite. If the dipolar interaction is weak in comparison with the
exchange energy $(g\mu_B)^2/a^3\ll |J|$ and if the argument of the summation is
approximated by its small {\bf q} limit one obtains
\begin{eqnarray}
T_N \sim {|J|\over ln{|J|\over E_0^\sigma}}~,
\end{eqnarray}
which coincides with the analogous formula derived by the Holstein-Primakoff
transformation (Eq. (18) in Ref.\cite{pich}). In the general case the
above sum [Eq. (18)] is evaluated with the full dispersion relation Eq. (6)
and by computing $100\times 100$ points in the 2D Brillouin zone and
determining the other points by linear extrapolation. The dipole sums have been
calculated via Ewald summation\cite{Bonsal77}.

Now we apply our theory to real quasi two-dimensional antiferromagnets.
Prominent examples of almost two-dimensional antiferromagnets are the
tetragonal antiferromagnetic halides K$_2$MnF$_4$, Rb$_2$MnF$_4$,
Rb$_2$MnCl$_4$ and (CH$_3$NH$_3$)$_2$MnCl$_4$. In these quadratic layer
structures the out-of-plane exchange interaction is neglegible in comparison to
the in-plane exchange interaction (about $10^{-4}$) \cite{Win731,Bir73} whereas
the dipole energy is larger by an order of magnitude. This two-dimensional
character has been shown experimentally by the absence of any dispersion along
the $z$-direction\cite{Bir73}. For these halides the measured exchange energy
$|J|$, the  lattice constant $a$, the energy gap $E^{\rm exp}_0$, the spin-flop
field $H^{\rm exp}_{\rm sf}$ and the transition temperature
$T_N^{\rm exp}$\cite{Bir73,Win,sch} are listed in table I. The spin-flop field
$H_{\bf sf}$ is the critical magnetic field at which the antiferromagnetic
N\'eel ground state changes to the spin-flop ground state. It can be calculated
by adding to the Hamiltonian, Eq. (1), the Zeeman energy\cite{footnote}.
{}From the full dispersion-relation the spin-flop field $H_{\bf sf}^\sigma$ is
defined by that field for which the magnon energy vanishes
\begin{equation}
H_{\bf sf}^\sigma = {1\over g\mu_B}E_0^\sigma,
\end{equation}
in close analogy to the formula obtained by the Holstein-Primakoff method (Eq.
(11) in Ref.\cite{pich}). The table also
contains the theoretical energy gaps $E^\sigma_0$ and $E_0$ calculated via Eq.
(15) and Eq. (9) of Ref.\cite{pich}, the resulting spin-flop fields
$H^\sigma_{\rm sf}$ and $H_{\bf sf}$ via Eq. (20) and Eq. (11) of
Ref.\cite{pich} and the theoretical transition temperature $T_N$ [Eq. (18)].
All these substances have spin $S=5/2$, which yields from Eq. (13) and (12)
$\sigma_0=2.30$ for pure isotropic antiferromagnets. This value is increased
only neglegibly by the dipolar interaction as can be seen from table I.

We find a good agreement with the measured the N\'eel temperature although our
theory accounts only for the dipolar interaction and no other anisotropy.
Corrections due to dipolar interactions between different planes
are neglegible because of the large lattice constant in $z$-direction; e.g.
the energy gap for K$_2$MnO$_4$ [Eq. (6)] is altered by
\begin{eqnarray}
E^\sigma_0(3D) = E^\sigma_0(2D)[1+{\cal O}(10^{-5})],
\end{eqnarray}
which justifies the application of the two-dimensional model. Note that the
nearest-neighbor exchange energy $J$ is the only parameter entering in our
theory.  Experimentally this parameter has been derived by fitting the measured
spin wave spectrum with a dispersion relation which is different of ours [Eq.
(6)].

For the halides listed in table I, the energy gap obtained from Eq. (15) and
the transition temperature are lower than the experimental values. The
following reasons may be responsible for that: (i) In the Holstein-Primakoff
approximation the softening of the magnons is neglected entirely. This leads to
an overestimate  of $T_N$. In the Callen method the magnons soften in the
entire Brillouin zone, thus particularly near the phase transition the
softening is overestimated and leads to a $T_N$ which is somewhat too low.
(ii) A small readjustment of $J$ could be necessary if our dispersion relation,
Eq. (6), is used to fit the data.
(iii) A small additional anisotropy from the crystal field might be present as
suggested by\cite{Win84}.

\acknowledgments This work has been supported by the German Federal Ministry
for Research and Technology (BMFT) under the contract number 03-SC2TUM.

\vskip 1cm
\begin{table}
\caption{Exchange energy $|J|$, lattice constant $a$, energy gap $E_0$,
spin-flop field $H_{\rm sf}$, N\'eel temperature $T_N$ and zero temperature
order parameter $\sigma_0$.}
\begin{tabular}{l|rc|ccc|ccc|cc|c}
& $|J|$ & $a $ & $E_0^{\rm exp} $ & $E_0^\sigma$ & $E_0$ & $H_{\rm sf}^{\rm
exp}$ & $H_{\rm sf}^\sigma$ &$H_{\rm sf}$ & $T_N^{\rm exp} $ & $T_N^{\rm th}$
& $\sigma_0$ \\
&[K]&[\AA]&[\rm K]&[K]& [K] & [T]&[T]&[T]&[K]&[K]&\\
\tableline
$K_2MnF_4$   & 8.5$^{\rm a}$ & 4.17$^{\rm a}$ & 7.4$^{\rm b}$ & 7.1 & 7.6 &
5.4$^{\rm d}$ & 5.3 & 5.6 & 42$^{\rm a}$ & 41 & 2.33 \\
$Rb_2MnF_4$  & 7.4$^{\rm c}$  & 4.20$^{\rm g}$ & 7.3$^{\rm b}$ & 6.5 & 7.0 &
& 4.8 & 5.1 & 38$^{\rm g}$  & 36 & 2.33 \\
$Rb_2MnCl_4$ & 11.2$^{\rm f}$ & 5.05$^{\rm e}$ & 7.5$^{\rm f}$ & 6.1 & 6.6 &&
4.5 & 4.9 & 56$^{\rm f}$ & 48 & 2.32 \\
$(CH_3NH_3)_2MnCl_4$ & 9.0$^{\rm f}$ & 5.13$^{\rm e}$ & & 5.3 & 5.7 &  & 3.9&
4.3 & 45$^{\rm f}$ & 39 & 2.32
\end{tabular}
\tablenotes{$^a$ Reference \cite{Bir73}}
\tablenotes{$^b$ Reference \cite{Win73}}
\tablenotes{$^c$ Reference \cite{Win731}}
\tablenotes{$^d$ Reference \cite{Win}}
\tablenotes{$^e$ Reference \cite{LB}}
\tablenotes{$^f$ Reference \cite{sch}}
\tablenotes{$^g$ Reference \cite{Bir70}}
\end{table}

\end{document}